\documentclass[12pt]{article}

\usepackage{fullpage} 
\newcommand{\E}{{\rm E}}
\newcommand{\Var}{{\rm Var}}
\newcommand{\Cov}{{\rm Cov}}
\newcommand{\Corr}{{\rm Corr}}

\newcommand \be  {\begin{equation}}
\newcommand \bea {\begin{eqnarray} \nonumber }
\newcommand \ee  {\end{equation}}
\newcommand \eea {\end{eqnarray}}

\begin{document}

\title{{\bf ``String'' formulation of the Dynamics of the Forward Interest Rate Curve}}
\author{\bf Didier Sornette}

\date{LPMC, CNRS UMR 6622 and Universit\'e des Sciences, Parc Valrose\\
 06108 Nice Cedex 2, France and\\
Institute of Geophysics and Planetary Physics, and
Department of Earth and Space Sciences, UCLA, Los Angeles, CA 90095-1567\\
sornette@naxos.unice.fr. \ \\ \ \\ }

\maketitle

\abstract{We propose a formulation of the term structure of interest rates in which
the forward curve is seen as the deformation of a string. We derive
the general condition that the partial differential equations governing the
motion of such string must obey in order to account for the condition 
of absence of arbitrage opportunities. This condition takes a form similar to 
a fluctuation-dissipation theorem, albeit on the same quantity (the forward rate),
 linking the bias to the covariance of variation 
fluctuations. We provide the general structure of the
models that obey this constraint in the framework of stochastic partial (possibly
non-linear) differential equations. We derive the general solution for the pricing
and hedging of interest rate derivatives within this framework, albeit
for the linear case (we also provide
in the appendix a simple and intuitive derivation of the standard European option
problem). We also show how the 
``string'' formulation simplifies into a standard $N$-factor model under 
a Galerkin approximation.

\vskip 2cm
Short title\,: Forward Interest Rate dynamics
\vskip 1cm
PACS: 02.50.-r    Probability theory, stochastic processes, and statistics

 05.40.+j  Fluctuation phenomena, random processes, and Brownian motion

 89.90.+n   Other areas of general interest to physicists

}

\thispagestyle{empty}
\newpage 
\pagenumbering{arabic}

\section{Introduction}

Imagine that Julie wants to invest \$1 for two years \cite{Brealey}. 
She can devise two possible strategies.
The first one is to put the money in a one-year bond at an interest rate $r_1$. At the 
end of the year, she must take her money and find another one-year bond, with interest
rate $r_2^{1}$ which is the interest rate in one year on a loan maturing in two years. The
final payoff of this strategy is simply $(1+r_1)(1+ r_2^{1})$.
The problem is that Julie cannot know for sure what will be
the one-period interest rate $r_2^{1}$ of next year. Thus, she can only estimate
a return by guessing the expectation of $r_2^{1}$.

Instead of making two separate investments of one year each, Julie could invest her money
today in a bond that pays off in two years with interest rate $r_2$. The final payoff is then
$(1+r_2)^2$. This second strategy is riskless as she knows for sure her return. Now, this strategy
can be reinterpreted along the line of the first strategy as follows. 
It consists in investing for one
year at the rate $r_1$ and for the second year at a {\it forward rate} $f_2$. The
forward rate is like the $r_2^{1}$ rate, with the essential difference that it is 
guaranteed\,: by buying the two-year bond, Julie can ``lock in'' an interest rate 
$f_2$ for the second year.

This simple example illustrates that the set of all possible bonds traded on the market
is equivalent to the so-called forward rate curve.
The forward rate $f(t,x)$ is thus the interest rate that can be contracted
at time $t$ for instantaneously riskless borrowing \footnote{``Instantaneous riskless''
describes the fact that the forward rate is the rate that applies for a small
time increment $\delta t$ as seen from equation (\ref{intfor}) below and is fixed
during this time, thus being locally riskless.} or lending at time $t+x$. 
It is thus a function or curve of the time-to-maturity $x$ \footnote{The maturity
of a financial product is simply its lifetime. In other
words, it is the time interval between the present and the time of extinction of 
the rights attached to the financial product.}, where $x$ plays the role
of a ``length'' variable, that deforms with time $t$. 
Its knowledge is completely equivalent to the set of bond prices $P(t,x)$ at
time $t$ that expire at time $t+x$ (see eq.(\ref{intfor}) below). 
The shape of the forward rate curve $f(t,x)$ incessantly fluctuates as a function of 
time $t$. These fluctuations are due to a combination of factors, including
 future expectation of the short-term interest rates, liquidity preferences,
 market segmentation and trading. It is obvious that the forward rate $f(t,x+\delta x)$ for 
 $\delta x$ small can not be very different from $f(t,x)$.
 It is thus tempting to see $f(t,x)$ as a ``string'' characterized by
a kind of tension which prevents too large local deformations that would not be financially
acceptable. This superficial analogy is in the follow up of the repetitious
intersections between finance and physics, starting with Bachelier \cite{Bachelier} who solved
the diffusion equation of Brownian motion as a model of stock market price fluctuations
five years before Einstein, continuing with 
the discovery of the relevance of L\'evy laws for cotton price
fluctuations by Mandelbrot \cite{Mandel} that can be compared with the present interest of such 
power laws for the description of physical and natural phenomena \cite{Dubrulle}.
We could go on and cite many other examples.
We investigate how to formalize mathematically this analogy between the forward rate curve
and a string. In this goal, we 
formulate the term structure of interest rates
as the solution of a stochastic partial differential 
equation (SPDE) \cite{Daprato}, following the 
physical analogy of a continuous curve (string) whose
shape moves stochastically through time. 

The equation of motion of macroscopic
physical strings is derived from conservation laws. The
fundamental equations of motion of
microscopic strings formulated to describe the fundamental particles \cite{Witten}
derive from global symmetry principles and dualities between long-range and short-range
descriptions. Are there similar principles that can guide the determination of the
equations of motion of the more down-to-earth financial forward
rate ``strings''? 

The situation is a priori much more difficult than in physics as illustrated by the
following picturial analogy quoted from the journalist N. Dunbar at
Financial Products magazine. Suppose that in the middle ages, 
before Copernicus and Galileo, the Earth
really was stationary at the centre of the universe, and only began moving
later on. Imagine that during the nineteenth century, when everyone
believed classical physics to be true, that it really was true, and quantum
phenomena were non-existent. 
These are not philosophical musings, but an attempt to portray how
physics might look if it actually behaved like the financial markets. Indeed,
the financial world is such that any insight is almost immediately used to trade for a
profit. As the insight spreads among traders, the ``universe'' changes
accordingly. As G. Soros has pointed out, market players are ``actors
observing their own deeds''. As E. Derman, head of quantitative strategies
at Goldman Sachs,  puts it, in physics you are
playing against God, who does not change his mind very often. In finance,
you are playing against God¹s creatures, whose feelings are ephemeral, at
best unstable, and the news on which they are based keep streaming in.
Value clearly derives from human beings, while mass, charge and
electromagnetism apparently do not. This has led to suggestions that a fruitful
framework to study finance and economy is to use evolutionary models inspired 
from biology and genetics.

This does not however guide us much for the determination of ``fundamental'' equations, 
if any. Here, we propose to use the condition of absence of arbitrage opportunity
\footnote{Arbitrage, also known as the Law of One Price, states that two
assets with identical attributes should sell for the same price and so should the same asset
trading in two different markets. If the prices differ, a profitable opportunity arises
to sell the asset where it is overpriced and to buy it where it is underpriced.}
and show that this leads to strong constraints on the structure of the governing equations
The basic idea is that, if there
are arbitrage opportunities (free lunches), they cannot live long or 
must be quite subtle, otherwise
traders would act on them and arbitrage them away. The no-arbitrage condition is 
an idealization of a self-consistent dynamical state of the market 
resulting from the incessant actions of the traders (arbitragers). It is not the out-of-fashion
equilibrium approximation sometimes described but rather
embodies a very subtle cooperative organization of the market.

We consider this condition as the fundamental
backbone for the theory. The idea to impose this requirement is not new and is in fact
the prerequisite of most models developed in the academic finance community. However, 
applying it in the present context is new.
Modigliani and Miller \cite{Modig,Miller} have indeed
emphasized the critical role played by arbitrage in determining the value of securities. 
It is sometimes suggested that transaction costs and other market imperfections
make irrelevant the no-arbitrage condition \cite{{BouSa}}. Let us address 
briefly this question before presenting our results.

Transaction costs in option replication and other hedging activities
\footnote{Finance is all about risks but some risks can be hedged, i.e. offset, 
by trading in different financial instruments.}
have been extensively investigated
since they (or other market ``imperfections'') clearly disturb
the risk-neutral argument and set option theory back a few decades.
Transaction costs induce, for obvious reasons, dynamic incompleteness, thus
preventing valuation as we know it since Black and Scholes \cite{Black}. 
However, the most efficient dynamic hedgers
(market makers) incur essentially no transaction
costs when owning options 
\footnote{In a nutshell, an option is an insurance for buying or selling.}
(see the appendix for a definition of options).
These specialized market makers compete with each other to
provide liquidity in option instruments, and maintain inventories in them.
They rationally limit their dynamic replication to their residual exposure, 
not their global exposure. In addition,
the fact that they do not hold options until maturity greatly reduces their
costs of dynamic hedging. They have an incentive in the acceleration of
financial intermediation. Furthermore, as options are rarely replicated
until maturity, the expected transaction costs of the short options depend
mostly on the dynamics of the order flow in the option markets -not on the
direct costs of transacting. The conclusion is that transaction costs are a
fraction of what has been assumed to be in the literature \cite{Taleb}. For the
efficient operators (and those operators only), markets are more
dynamically complete than anticipated. This is not true for a second
category of traders, those who
merely purchase or sell financial instruments that are subjected to dynamic
hedging. They, accordingly, neither are equipped for dynamic hedging, nor
have the need for it, thanks to the existence of specialized and more
efficient market makers. The examination of their transaction costs in the
event of their decision to dynamically replicate their options is of no
true theoretical contribution.

A second important point is that the existence of transaction costs should not
be invoked as an excuse for disregarding the no-arbitrage condition but rather
should be constructively invoked to study its impacts on the models.

This work expands our previous work \cite{Santa}
which introduced stochastic strings multiplied by volatility functions to shock forward rates. 
Our present approach directly deals with the SPDE equation for the forward rates in contrast
to the shocks that drive the forward rate curve that were treated in Ref.\cite{Santa}.
Our model provides a further extension of the term structure model of Heath, 
Jarrow and Morton \cite{HJM} and is as
parsimonious and tractable as the traditional HJM model, but is capable of
generating a much richer class of dynamics and shapes of the forward rate
curve. Its main motivation is to address the interplay between 
external factors represented by stochastic components (noise) and possible non-linear
dynamics. 


\section{Definitions}


We postulate the existence of a stochastic discount factor (SDF) that
prices all assets in this economy and denote it by $M$. This process is also
termed the pricing kernel, the pricing operator, or the state price density. We use
these terms interchangeably. Ref.\cite{Duffie} is an excellent reference for the
theory behind the SDF. It is well known that assuming that no dynamic
arbitrage trading strategies can be implemented by trading in the financial
securities issued in the economy is roughly
equivalent to the existence of a
strictly positive SDF. For no arbitrage opportunities to exist, the product of
$M$ with the value process of any investment strategy must be a martingale
\footnote{Technically, recall that a martingale is a family of random variables $\xi(t)$ such that
the mathematical expectation of the increment 
$\xi(t_2) - \xi(t_1)$ (for arbitrary $t_1 < t_2$), conditioned on the past values 
$\xi(s)$ ($s \leq t_1$), is zero. The drift of a martingale is thus zero.
This is different from the Markov process, which is better known
in the physical community, defined by the independence of the next increment
on past values.}.
Under an adequate definition of the space of admissible trading strategies, the
product $MV$ is a martingale, where $V$ is the value process of any admissible
self-financing trading strategy implemented by trading on financial securities. Then,
\be 
V(t)=\E_t \left[V(s) \frac{M(s)}{M(t)} \right]~, \label{martcond} 
\ee 
where $s$ is a future date and
$\E_t \left[x\right]$ denotes the mathematical expectation of $x$ taken at time
$t$. In particular, we require that a bank account and zero-coupon discount
bonds of all maturities satisfy this condition.

A security is referred to as a (floating-rate) bank account, if it is ``locally
riskless'' \footnote{A security is ``locally riskless'' if, over an
instantaneous  time interval, its value varies deterministically. It may
still be random, but there is no Brownian  term in its dynamics.}. Thus, the value at
time $t$, of an initial investment of $B(0)$ units in the bank account that is
continuously reinvested, is given by the following process 
\be 
B(t) = B(0)\exp \left\{ \int_{0}^{t}r(s)ds \right\}~, 
\ee 
where $r(t)$ is the instantaneous nominal interest rate.

We further assume that, at any time $t$, riskless discount bonds of all {\it
maturity dates} $s$ trade in this economy and let $P(t,s)$ denote the time $t$
price of the $s$ maturity bond. We require that $P(s,s)=1$, that $P(t,s)>0$ and
that $\partial P(t,s)/\partial s$ exists.

Instantaneous forward rates at time $t$ for all {\it times-to-maturity} $x>0$,
$f(t,x)$, are defined by 
\be
 f(t,x) = -  \frac{\partial \log P(t,t+x)}{\partial x}~,  
 \ee
which is the rate that can be contracted at time $t$ for instantaneously
riskless borrowing or lending at time $t+x$. We require that the initial
forward curve $f(0,x)$, for all $x$, be continuous.

Equivalently, from the knowledge of the instantaneous forward rates for all
times-to-maturity between $0$ and time $s-t$, the price at time $t$ of a bond
with maturity $s$ can be obtained by 
\be 
P(t,s) = \exp \left\{ - \int_0^{s-t}
f(t,x) dx  \right\} ~. \label{intfor} 
\ee 
Forward rates thus fully represent the
information in the prices of all zero-coupon bonds.

The spot interest rate at time $t$, $r(t)$, is the instantaneous forward
rate at time $t$ with time-to-maturity $0$, 
\be 
r(t) = f(t,0) ~. \label{rate} 
\ee

For convenience, we model the dynamics of forward rates. Clearly, we could as
well model the dynamics of bond prices directly, or even the dynamics of the
yields to maturity of the zero-coupon bonds.
We use forward rates with fixed {\it time-to-maturity} rather than fixed {\it
maturity date}. The model of HJM starts from processes for forward
rates with a fixed maturity date. This is different from what we do. If we use a
``hat'' to denote the forward rates modeled by HJM, 
\be
 \hat{f}(t,s)=f(t,s-t) 
\ee
or, equivalently, 
\be
 f(t,x)=\hat{f}(t,t+x) 
 \ee
 for fixed $s$. Musiela (1993) and
Brace and Musiela \cite{BraceMu} define forward rates in the same fashion. Miltersen,
Sandmann and Sondermann \cite{Milter}, and Brace, Gatarek and Musiela \cite{BraceGa} use
definitions of forward rates similar to ours, albeit for non-instantaneous
forward rates. Modelling forward rates with fixed time-to-maturity is more
natural for thinking of the dynamics of the entire forward curve as the
shape of a string evolving in time. In contrast, in HJM, forward rate processes
disappear as time reaches their maturities. Note, however, that we still impose the
martingale condition on bonds with fixed {\it maturity date}, since these
are the financial instruments that are actually traded.


\section{Stochastic strings as solutions of SPDE's}


In a nutshell, the contribution of this paper consists in modelling
the dynamical evolution of the forward rate curve by stochastic {\it partial} 
differential equations (SPDE's) \cite{Daprato}.
In the context of continuous-time finance, 
this is the most natural and general extension that can be
performed \footnote{Further extensions will include fractional differential
equations and integro-differential equations, including jump processes.}.

Financial and economic time series are often described to a first degree of 
approximation as random walks, following the precursory work of Bachelier \cite{Bachelier} and
Samuelson \cite{Samuelson}. A random walk is the mathematical translation of the trajectory
followed by a particle subjected to random velocity variations. The analogous physical system
described by SPDE's is a {\it stochastic string}. The length along the
string is the time-to-maturity and the string configuration (its transverse deformation)
gives the value of the forward rate $f(t,x)$ at a given time for each
time-to-maturity $x$. The
set of admissible dynamics of the configuration of the  string as a function of
time depends on the structure of the SPDE. Let us for the time being 
restrict our attention to SPDE's in which the highest derivative is second order. This
second order derivative has a simple physical interpretation\,: the string is
subjected to a tension, like a piano chord, that tends to bring it back to zero
transverse deformation. This tension forces the ``coupling'' among different
times-to-maturity so that the forward rate curve is at least continuous. In
principle, the most general formulation would consider SPDE's with terms of
arbitrary derivative orders.\footnote{Higher order derivatives also have an
intuitive physical interpretation. For instance, going up to fourth order
derivatives in the SPDE correspond to the dynamics of a {\it beam}, which has
bending elastic modulus tending to restore the beam back to zero deformation,
even in absence of tension.} However, it is easy to show that the tension
term is the dominating restoring force, when present, for deformations of the string
(forward rate curve) at long ``wavelengths'', i.e. for slow variations
along the time-to-maturity axis. Second order SPDE's are thus generic in the sense of a
systematic expansion \footnote{There are situations where the tension can
be made to vanish (for instance in the presence of a rotational symmetry) and then the
leading term in the SPDE becomes the fourth order ``beam'' term.}.

In the framework of second order SPDE's, we consider {\it hyperbolic}, {\it
parabolic} and {\it elliptic} SPDE's, to characterize the dynamics of the
string along two directions~: inertia or mass, and viscosity or subjection to drag
forces. A string that has ``inertia'' or, equivalently, ``mass'' per unit
length, along with the tension that keeps it continuous, is characterized by the
class of {\it hyperbolic} SPDE's. For these SPDE's, the highest order derivative in time
has the same order as the highest order derivative in distance along the string
(time-to-maturity). As a consequence, hyperbolic SPDE's present wave-like
solutions, that can propagate as pulses with a ``velocity''.  In this class, we
find the so-called ``Brownian sheet'' which is the direct
generalization of Brownian motion to higher dimensions, that preserves continuity in
time-to-maturity. The Brownian sheet is the surface spanned by the string
configurations as time goes on. The Brownian sheet is however
non-homogeneous in time-to-maturity, which led us to examine 
other processes.

If the string has no
inertia,\footnote{Or if the inertia term is negligible compared to the drag
term proportional to the first time derivative (so-called overdamped dynamics).} its
dynamics are characterized by {\it parabolic} SPDE's. These
stochastic processes lead to smoother diffusion of shocks through time, along time-to-maturity. 

Finally, we mention the third class of SPDE's of second-order,
namely elliptic partial differential equations. Elliptic SPDE's give processes
that are differentiable both in $x$ and $t$. Therefore, in the strict limit of
continuous trading, these stochastic processes correspond to locally riskless interest
rates.

For the sake of completeness and clarity, we briefly summarize useful facts
about PDE's (See for instance Ref.\cite{Morse} and, in particular, their
classification and the intuitive meaning behind it. We restrict our
discussion to
two-dimensional examples. Their general form reads 
\be 
A(t,x) {\partial^2 f(t,x)
\over \partial t^2} + 2 B(t,x) {\partial^2 f(t,x) \over \partial t \partial x} +
C(t,x) {\partial^2 f(t,x) \over \partial x^2} = F(t,x, f(t,x), {\partial f(t,x) \over
\partial t}, {\partial f(t,x) \over \partial x}, S) ~, 
\label{zz1} 
\ee 
where $f(t,x)$ is the forward rate curve. $S(t,x)$ is the ``source'' term 
that will be generally taken to
be Gaussian white noise $\eta (t,x)$ characterized by the covariance
\be
\Cov \left[ \eta (t,x),~ \eta (t',x') \right] = \delta (t-t') ~\delta (x-x')~,
\label{cvgherfv}
\ee
where $\delta$ denotes the Dirac distribution.
Expression (\ref{zz1}) is the most general
second-order SPDE in two variables. 
For arbitrary non-linear terms in $F$, 
the existence of solutions is not warranted and a case by case study must be performed.
For the cases where $F$ is linear, the solution $f(t,x)$ exists
and its uniqueness
is warranted once ``boundary'' conditions are given, such as, for instance, the
initial value of the function $f(0,x)$ as well as any constraints on
the particular form of equation (\ref{zz1}). 

Equation (\ref{zz1}) is defined by its {\it characteristics}, which are
curves in the $(t,x)$ plane that come in two families  of equation\,: 
\be A dt = (B + \sqrt{B^2 - AC}) dx~, 
\ee 
\be 
A dt = (B - \sqrt{B^2 - AC}) dx~. 
\ee 
These characteristics are the geometrical loci of the propagation of the boundary
conditions.

Three cases must be considered.

\noindent $\bullet$ When $B^2 > AC$, the characteristics are real curves and the corresponding
SPDE's are called ``hyperbolic''. For such hyperbolic SPDE's, the natural
coordinate system is formed from the two families of characteristics. Expressing
(\ref{zz1}) in terms of these two natural coordinates $\lambda$ and $\mu$,
we get the ``normal form'' of hyperbolic SPDE's\,:
\be
{\partial^2 f \over \partial \lambda \partial \mu} =
P(\lambda,\mu) {\partial f \over \partial \lambda} + Q(\lambda,\mu)
{\partial f \over \partial \mu} + R(\lambda,\mu) f + S(\lambda,\mu)~.
\ee
The special case $P=Q=R=0$ with
$S(\lambda,\mu)=\eta(\lambda,\mu)$
corresponds to the so-called Brownian sheet, well studied in the mathematical
literature as the 2D {\it continuous} generalization of the Brownian motion.

\noindent $\bullet$ When $B^2 = AC$, there is only one family of characteristics, of equation
\be
A dt = B dx~.
\ee
Expressing (\ref{zz1}) in terms of the natural characteristic coordinate
$\lambda$ and keeping $x$, we get the
``normal form'' of parabolic SPDE's\,:
\be
{\partial^2 f \over \partial x^2} =
K(\lambda,\mu) {\partial f \over \partial \lambda} + L(\lambda,\mu)
{\partial f \over \partial x} + M(\lambda,\mu) f + S(\lambda,\mu)~.
\ee
The diffusion equation, well-known to be associated to the Black-Scholes
option pricing model, is of this type. The main difference with the hyperbolic
equations
is that it is no more invariant with respect to time-reversal $t \to -t$.
Intuitively, this is due to the fact that the diffusion equation is not conservative,
the information content (negentropy) continually decreases as time goes on.

\noindent $\bullet$ When $B^2 < AC$, the characteristics are {\it not} real curves and the
corresponding
SPDE's are called ``elliptic''. The equations for the
characteristics are complex
conjugates of each other and we can get the ``normal form'' of elliptic
SPDE's by using the real and imaginary parts of these complex coordinates $z=u \pm iv$\,:
\be
{\partial^2 f \over \partial u^2} + {\partial^2 f \over \partial v^2} =
T {\partial f \over \partial u} + U
{\partial f \over \partial v} + V f + S~.
\ee
There is a deep connection between the solution of elliptic SPDE's and
analytic functions of complex variables. 

We have shown \cite{Santa}
 that hyperbolic and parabolic SPDE's provide processes
reducing locally to standard Brownian motion at fixed time-to-maturity,
while elliptic
SPDE's give locally riskless time evolutions. Basically, this stems from
the fact that
the ``normal forms'' of second-order hyperbolic and parabolic SPDE's
involve a first-order
derivative in time, thus ensuring that the stochastic processes are locally
Brownian in
time. In contrast, the ``normal form'' of second-order elliptic SPDE's involve
a second-order derivative with respect to time, which is the cause for the
differentiability
of the process with respect to time. Any higher order SPDE will be
Brownian-like in time if
it remains of order one in its time derivatives (and higher-order in the
derivatives with
respect to $x$).


\section{No-arbitrage condition\,: derivation of the general condition}


We now proceed to derive the general condition that the forward rate equation must
obey to be compatible with the no-arbitrage constraint.

From (\ref{intfor}), we get
\be
d_t \log P(t,s) = f(t,x)~dt  - \int_0^{x} dy ~d_t f(t,y)~,
\label{erdxwwxx}
\ee
where $x \equiv s-t$. 
We need the expression of ${d P(t,s) \over P(t,s)}$ which is obtained from
(\ref{erdxwwxx}) using Ito's calculus \cite{Ito}.
In order to get Ito's term in the drift,
recall that it results from the fact that, if $f$ is stochastic, then
\be
d_t F(f) = {\partial F \over
df} d_t f + {1 \over 2} \int dx \int dx' {\partial^2 F \over \partial f(t,x) 
\partial f(t,x')} \Cov \left[ d_t f(t,x),d_t f(t,x') \right]~,
\ee
where $\Cov \left[ d_t f(t,x),d_t f(t,x') \right]$ is the covariance of the
time increments of $f(t,x)$.

Using this Ito's calculus, we obtain
$$
{dP(t,s) \over P(t,s)} = \biggl[ dt~f(t,x) - \int_0^{x} dy ~E_{t,d_t f}(t,y)
+ {1 \over 2} \int_0^{x} dy  \int_0^{x} dy' ~\Cov \left[ d_t f(t,y) d_t f(t,y') \right] \biggl]
$$
\be
- \int_0^{x} dy ~[d_t f(t,y) - E_{t,d_t f}(t,y)]~.
\ee
We have explicitely taken into account the fact that 
$d_t f(t,x)$ may have in general a non-zero drift, i.e. its
expectation 
\be
E_{t,d_t f}(t,x) \equiv \E_t \left[ d_t f(t,x) | f(t,x)\right]
\ee
conditioned on $f(t,x)$ is non-zero. 

 The no-arbitrage condition for buying and holding bonds implies that $PM$ is a martingale
 in time, for any bond price $P$. Technically this amounts to imposing that the drift of
 $PM$ be zero\,:
\be 
f(t, x) = f(t,0) + \int_0^{x} dy ~{E_{t,d_t f}(t,y) \over dt}
- {1 \over 2} \int_0^{x} dy  \int_0^{x} dy' ~
c(t,y,y') + o(1)~,
\label{rfdfsd}
\ee
assuming that $d_t f(t,x)$ is not correlated with the stochastic process driving the pricing 
kernel and using the definitions
\be
c(t,y,y') dt = \Cov  [d_t f(t,y) d_t f(t,y')] ~,
\ee
and $r(t) = f(t,0)$ as given by (\ref{rate}).
In (\ref{rfdfsd}), the notation $o(1)$ designs terms of order $dt$ taken to a positive
power.

Expression (\ref{rfdfsd}) is the fundamental constraint that a SPDE for $f(t,x)$ must 
satisfy in order to obey the no-arbitrage requirement. As in other formulations,
this condition relates the drift to the volatility.

It is useful to parametrize, without loss of generality, 
\be
{E_{t,d_t f}(t,x) \over dt} = {\partial f(t,x) \over \partial x} + h(t,x)~,
\label{erfdffdf}
\ee
where $h(t,x)$ is a priori arbitrary. The usefulness of this parametrization
(\ref{erfdffdf}) stems from the fact that it allows us to get rid of the terms
$f(t, x)$ and $f(t,0)$ in (\ref{rfdfsd}). Indeed, they cancel out with the integral over $y$
of ${E_{t,d_t f}(t,y) \over dt}$. Taking the derivative with respect to $x$ of
the no-arbitrage condition (\ref{rfdfsd}), we obtain
\be
h(t,x) = - {1 \over 2} \int_0^x ~[c(t,y,x) + c(t,x,y)]~.
\ee
In sum, we have the following constraint that the SPDE for $f(t,x)$ must satisfy
\be
\E_t \left[ d_t f(t,x) | f(t,x)\right] = dt~{\partial f(t,x) \over \partial x} 
-  {1 \over 2} \int_0^x dy ~\biggl(\Cov  [d_t f(t,y) d_t f(t,x)] + \Cov  [d_t f(t,x) d_t f(t,y)]\biggl) ~,
\label{erpoiuyy}
\ee
which, for symmetric covariance, lead to 
\be
\E_t \left[ d_t f(t,x) | f(t,x)\right] = dt~{\partial f(t,x) \over \partial x} 
-  \int_0^x dy ~\Cov [d_t f(t,y) d_t f(t,x)] ~.
\label{terfdvxc}
\ee

This expression (\ref{terfdvxc}) is reminiscent of the fluctuation-dissipation 
theorem in the Langevin formulation of the Brownian motion \cite{Strato},
linking the  drag coefficient (analog to the l.h.s.) to the integral over time
of the correlation function of the fluctuation forces (analog of the second term
\footnote{The first term of the r.h.s. is a drift term that disappear by
a galilean transformation of frame.} of the r.h.s.).
In the usual fluctuation-dissipation theorem, the drag coefficient is 
determined self-consistently as a function of the amplitude and correlation of the
fluctuating force in order to be compatible with the equilibrium distribution. 
Similarly, here the no-arbitrage condition determines self-consistently the
conditional drift from the covariance of the fluctuations.
In contrast, notice however that the same quantity $f$ enters in both sides of (\ref{erpoiuyy})
and (\ref{terfdvxc}). Thus, the no-arbitrage condition imposes a condition on the structure
of the partial differential equations that govern the dynamical evolution of the forward
rates $f(t,x)$.


\section{General structure of the SPDE's compatible with the no-arbitrage condition}


To derive the structure of the SPDE's compatible with the no-arbitrage condition
(\ref{erpoiuyy}), we come back to the formulation of \cite{Santa}
and parametrize the time increment of the forward rate as
\be
d_t f(t,x) = \alpha(t,x) dt + \sigma(t,x) d_t Z(t,x)~,
\label{modelsheet}
\ee
where $Z(t,x)$ is a infinite dimensional stochastic process which is continuous in
$x$ and $t$ and $\alpha(t,x)$ and $\sigma(t,x)$ are a priori arbitrary functions of 
$f(t,x)$.

Using this formulation (\ref{modelsheet}), we have shown previously \cite{Santa}
 that the condition of no arbitrage 
leads to the following dynamics for the forward rates
\be
d_t f(t,x) =  dt \biggl( \frac{\partial f(t,x)}{\partial x} + A(t,x) \biggl) 
+ \sigma(t,x) ~d_t Z(t,x)~,
\label{bphhbcgg}
\ee
where 
\be
A(t,x) = \sigma(t,x) \biggl( \int_0^x dy~\sigma(t,y) c_Z(t,y,x) \biggl)~,
\label{aaaaqqzzzs}
\ee
and
\be
c_Z(t,y,y') ~dt \equiv \Cov \left[ d_t Z(t,y),d_t Z(t,y') \right]~.
\label{coroduu}
\ee
We stress that, by construction, the form (\ref{bphhbcgg}) with 
(\ref{aaaaqqzzzs}) and (\ref{coroduu})
automatically satisfies the condition (\ref{erpoiuyy}). 

In order to get the form of the allowed SPDE's for $f(t,x)$, 
we recall the requirements that are convenient to impose $Z$,
without loss of generality.
\begin{enumerate}
\item $Z(t,x)$ is continuous in $x$ at all times $t$;
\item $Z(t,x)$ is continuous in $t$ for all $x$;
\item $Z(t,x)$ is a martingale in time $t$, $\E \left[ d_t Z(t,x) \right]=0$,
for all $x$;
\item The variance of the
increments, $\Var \left[ d_t Z(t,x) \right]$, does not depend on $t$ or $x$;
\item The correlation of the increments,  $\Corr \left[ d_t Z(t,x),  d_t
Z(t,x')  \right]$, does not depend on $t$.
\end{enumerate}

Then, a fairly general stochastic function $Z(t,x)$ obeying these requirements is such that
(Santa-Clara and Sornette, 1997)
\be
d_t Z(t,x) = dt~ {1 \over \sqrt{j(x)}} \int_0^{j(x)} dy~\eta(t,y)~,
\label{cxcvvbx}
\ee
where $\eta (t,x)$ is a Gaussian white noise characterized by the covariance
(\ref{cvgherfv}). This leads to the following
covariance function for $Z(t,x)$\,:
\be
c_Z(t,y,y') = \sqrt{j(x) \over j(x')}  ~~~~{\rm for~~} j(x) < j(x')~,
\label{dfcghj}
\ee
and the roles of $x$ and $x'$ in (\ref{dfcghj}) are reversed if $j(x) > j(x')$.

Inserting (\ref{cxcvvbx}) in (\ref{bphhbcgg}), we get
\be
\frac{\partial f(t,x)}{\partial t} - \frac{\partial f(t,x)}{\partial x}
= \sigma(t,x) \biggl( \int_0^x dy~\sigma(t,y) \sqrt{j(y) \over j(x)}  \biggl)
+ {\sigma(t,x) \over \sqrt{j(x)}} \int_0^{j(x)} dy~\eta(t,y)~.
\ee
Multiplying both sides of this equation by ${\sqrt{j(x)} \over \sigma(t,x)}$ and taking
the partial derivative with respect to $x$ finally yields
\be
{\partial \over \partial x} \biggl[ {\sqrt{j(x)} \over \sigma(t,x)}
\biggl(\frac{\partial f(t,x)}{\partial t} - \frac{\partial f(t,x)}{\partial x}\biggl) \biggl]
= {\partial \over \partial x} \biggl[ \sqrt{j(x)} \int_0^x dy~\sigma(t,y) \sqrt{j(y) \over j(x)} 
\biggl] + \sqrt{|{dj(x) \over dx}|} ~\eta(t,x)~,
\label{retfdfund}
\ee
where we have used that $\eta(t,j(x)) = \eta(t,x) / \sqrt{|dj(x)/dx|}$.
This provides a first class of SPDE, which is in general non-linear since the volatility
$\sigma(t,x)$ can be an arbitrary function of $f(t,x)$.

An even more general class of SPDE's for $f(t,x)$ is obtained by using
the most general stochastic function $Z(t,x)$ obeying the requirements $1-5$ 
(Santa-Clara and Sornette, 1997)\,:
\be
d_t Z(t,x) = dt~ {1 \over \sqrt{l(x)}} \int_0^{j(x)} dy ~
\sqrt{{d \over dy}l\biggl([j]^{-1}\biggl)(y)}~ \eta(t,y)~.
\label{fcvgfdxc}
\ee
The correlation of the increments is
\be
c_Z(t,x,x') = \sqrt{l(x) \over l(x')} ~, ~~~~{\rm if ~~} j(x) < j(x')~.
\label{rfdscxwwwww}
\ee
The role of $x$ and $x'$ in (\ref{rfdscxwwwww}) are
inverted
if $j(x) > j(x')$. This provides a 
generalization to (\ref{cxcvvbx}) since a different function appears in the
correlation function and in the inequality condition on $x$ and $x'$.

Inserting (\ref{fcvgfdxc}) in (\ref{bphhbcgg}), we get
\be
\frac{\partial f(t,x)}{\partial t} - \frac{\partial f(t,x)}{\partial x}
= \sigma(t,x) \biggl( \int_0^x dy~\sigma(t,y) \sqrt{l(y) \over l(x)}  \biggl)
+ {\sigma(t,x) \over \sqrt{l(x)}} \int_0^{j(x)} dy~
\sqrt{{d \over dy}l\biggl([j]^{-1}\biggl)(y)}~\eta(t,y) \biggl)~.
\ee
Multiplying both sides of this equation by ${\sqrt{l(x)} \over \sigma(t,x)}$ and taking
the partial derivative with respect to $x$ finally yields
\be
{\partial \over \partial x} \biggl[ {\sqrt{l(x)} \over \sigma(t,x)}
\biggl(\frac{\partial f(t,x)}{\partial t} - \frac{\partial f(t,x)}{\partial x}\biggl) \biggl]
= {\partial \over \partial x} \biggl[ \sqrt{j(x)} \int_0^x dy~\sigma(t,y) \sqrt{l(y) \over l(x)} 
\biggl] + \sqrt{d l(x) \over dx} \sqrt{|{dj(x) \over dx}|} ~\eta(t,x)~.
\label{retfdfqsdqund}
\ee
This provides the most general class of SPDE describing the dynamical evolution of the 
forward rates $f(t,x)$ due to the interplay of a stochastic forcing $\eta$ and non-linearities
embodied in $\sigma(t,x)$. 

According to the classification briefly summarized above, these equations (\ref{retfdfund})
and (\ref{retfdfqsdqund}) are both of the hyperbolic class. It is noteworthy that the 
no-arbitrage condition excludes the parabolic class. Physical strings obey hyperbolic
equations for zero or small dissipation and parabolic equations in the over-damped limit.
A posteriori, it is not surprising that we find that 
the no-arbitrage condition, which implies the 
absence of market friction, corresponds to the first class. Notice that 
the term $\frac{\partial f(t,x)}{\partial t} - \frac{\partial f(t,x)}{\partial x}$ can
be replaced by $\frac{\partial f(t,x)}{\partial t}$ when changing the variables $(t,x)$
to $(t, s=t+x)$ and the l.h.s. of (\ref{retfdfqsdqund}) become of the form
${\partial \over \partial s} \biggl[ {\sqrt{l(s)} \over \sigma(t,s)}
\frac{\partial {\hat f}(t,s)}{\partial t}  \biggl]$.

These equations (\ref{retfdfund}) and (\ref{retfdfqsdqund}) are characterized by
 partial derivatives and two source terms in the r.h.s., the first one being locally
 adapted and the second one corresponding to the influence of external noise. 
 These equations can be explicitely solved if $\sigma$ is specified and is independent
 of $f(t,x)$, thus keeping the equations linear, as done in (Santa-Clara and Sornette, 1997).
The problem of finding a solution of these equations when $\sigma(t,x)$ depends on $f(t,x)$
is much more complex and belongs to the vast class of generally unsolved non-linear
partial differential equations. Such equations have been encountered in many different
areas, in particular some apparently particularly simple nonlinear PDE's have been found
to exhibit the most complex phenomenology one can imagine. A vivid example is the Navier-Stokes
equation of fluid motion leading to fluid turbulence when its single parameter (viscosity) goes
to zero \cite{Frisch}. Extrapolating on these superficial analogies, 
we conjecture that it might be
possible to postulate a simple form for the nonlinear dependence of $\sigma$ as a function of
$f(t,x)$, such that one or two real parameters might embody the full phenomenology of observed
forward rate statistics.


\section{A parametric example}


Empirical observation shows that
correlation between
two forward rates, with maturities separated by a given interval, increases
with maturity. We have shown previously \cite{Santa} that
the parametrization $j(x) = e^{i(x)}$ leads to the simple and general
condition that $i(x)$ must be
a function which increases more slowly than $x$, i.e. $i(x)$ must be concave (downward).
In order words, $j(x)$ must grow slower than an exponential. 
Any expression like
$j(x) = \exp(kappa x^alpha)$ with an exponent $0 < \alpha < 1$ qualifies. This
corresponds to
\be c_Z(t,x,y) = e^{-\kappa |x^{\alpha}-y^{\alpha}|} ~.
\label{corralpha} 
\ee

Reporting this choice in (\ref{retfdfund}) yields the following SPDE\,:
$$
{\partial \over \partial x} \biggl[ {e^{(\kappa/2) x^{\alpha}} \over \sigma(t,x)}
\biggl(\frac{\partial f(t,x)}{\partial t} - \frac{\partial f(t,x)}{\partial x}\biggl) \biggl]
= 
$$
\be
{\partial \over \partial x} \biggl[ 
 \int_0^x dy~\sigma(t,y) e^{(\kappa/2) y^{\alpha}} 
\biggl] + \sqrt{\kappa \alpha} x^{\alpha -1 \over 2} e^{(\kappa/2) x^{\alpha}}
 ~\eta(t,x)~.
\label{retfdqgfgqsfund}
\ee


\section{Reduction to $N$-factors models}


Let us now show how our model naturally encompasses the usual HJM formulation
of forward interest rates
in terms of $N$ factors. The idea is that $N$ factors models can be obtained as truncations
at the order $N$ in a way similar to a galerkin approximation, corresponding to a 
 finite ``resolution'', of an infinite expansion over the eigenfunctions of the 
operator defining the partial differential equation. This point of view allows one
to clearly see both the relationship with previous formulations and their limitations.
It also gives a justification to $N$ factor models in the following sense. As the 
stochastic string description constitutes the most general description of
the forward rate curve,
the fact that the $N$ factors models emerge naturally by a truncation of this general formulation,
amounting to limit the resolution of the fluctuations in the time of maturity axis,
justifies their mathematical status as simply a degree of approximation of an ideal
general description.

To keep the exposition as general as possible, let us call ${\cal L}$ the {\it linear}
operator defined by
\be
{\cal L} f(t,x) = \eta(t,x)~.
\label{operiardjdk}
\ee
We simplify the problem by restricting our analysis to linear SPDE's. This allows us
to use the general property of linear operators that the Green function $G(t,x|v,y)$ defined by
\be
{\cal L} G(t,x|v,y) = \delta(t-v) ~\delta(x-y)
\label{operiardjdk2}
\ee
can be expressed as \cite{Morse}
\be
G(t,x|v,y) = \sum_{n=1}^{\infty} l_n(v,t) ~\psi_n(x)~\phi_n(y)~.
\label{tfgvcbb}
\ee
The $\psi_n(x)$ are the eigenfunctions of ${\cal L}$. The $\phi_n(x)$ are the 
eigenfunctions of the adjoint operator ${\cal L}^*$, which is the same as ${\cal L}$ for
the hyperbolic string wave equations considered above (self-adjoint case) (but would not
be the same for a parabolic string equation). The $l_n(v,t)$ are a set of functions 
which depend on ${\cal L}$. Now, the eigenfunctions of ${\cal L}$ form an orthogonal 
basis on which one can expand the source term $\eta(t,x)$\,:
\be
\eta(t,x) = \sum_{j=1}^{\infty} \eta_j(t)~\psi_j(x)~.
\ee
The delta-covariance property of $\eta(t,x)$ allows us to choose that of the $\eta_j(t)$
as follows\,:
\be
\Cov[\eta_j(t),\eta_k(t')] = g_{jk}(t)~\delta(t-t')~,
\label{tfdccvbwlkkjlf}
\ee
where $g_{jk}(t)$ depends on the form of the operator.
Using the general solution of (\ref{operiardjdk}) which reads
\be
f(t,x) = f(0,x) + \int_0^t dv \int_{-\infty}^{\infty} dy
~G(t,x|v,y)~ \eta(v,y)~,
\label{dggvvgggg}
\ee
we obtain
\be
f(x,t) =  \int_0^t dv \int_{-\infty}^{\infty} dy
\sum_{n=1}^{\infty} l_n(v,t) ~\psi_n(x)~\phi_n(y)
\sum_{j=1}^{\infty} \eta_j(v)~\phi_j(y)~.
\label{dggvvgggdfdfg}
\ee
By the orthogonality condition $\int_{-\infty}^{\infty} dy
\phi_n(y)~\psi_j(y) = \delta_{nj}$, we obtain
\be
f(x,t) =  \int_0^t dv 
\sum_{n=1}^{\infty} l_n(v,t) ~\psi_n(x) \eta_n(v)~.
\label{uuuuvgggdfdfg}
\ee
We recognize a $N$-factor model by truncating this sum at $n=N$ and
with the identification
\be
\sqrt{g_{nn}(t)} ~l_n(v,t)~\psi_n(x)  \equiv \sigma_n(t,x)~,
\ee
and
\be
{\eta_n(t) dt \over \sqrt{g_{nn}(t)}} \equiv d_t W_n(t) ~.
\ee
Since the functions $\psi_n(x)$ are more and more complicated or convoluted
as the order $n$ increases, the truncation at a finite order $N$ indeed
corresponds to a finite resolution in the driving of the forward rate curve.
For instance, $\psi_1(x)$ can be a constant, $\psi_2(x)$ can be a parabola
with a single maximum, $\psi_3(x)$ can be a quartic (two up maxima and one down maximum), etc.


\section{Option Pricing and Replication}


In this section, we derive the general equation for 
 the pricing and hedging of interest rate
derivatives. Our derivation is restricted to linear SPDE's, i.e. to the
cases where $\sigma(t,x)$ is not an explicit function of the forward rates $f$.

\subsection{The general case}

In general, European term structure derivatives in our model have a payoff
function of the form
\be
C(t,\{P(s)\},r(t)) ~,
\label{call}
\ee
where the argument $\{P(s)\}$ indicates that the payoff depends on the price at
time $s$ of all
bonds of different time of maturities, i.e. on the
full forward rate curve.\footnote{Expressing the payoff as a function of
the full forward curve is the same as expressing it as a function of the
corresponding continuous set of bonds.} The problem we address is that of
hedging this claim by trading bonds. Hedging these claims in general
implies trading in an infinity of
bonds, so that the claim's price at time $t$ will be a function of the
entire forward rate
curve.

There is thus in general no hope of being able
to perfectly hedge interest rate contingent claims with a finite number of
bonds. We therefore extend the
space of admissible trading strategies to include density valued portofolios.

We denote by $h(t,s)$ the density (number) of bonds held in the portfolio
at time $t$, with maturity $s$, and let $g(t)$ be the amount invested
in the bank account. Then, the value of an investment strategy is
$$
V(t) = g(t)B(t) + \int_t^{\infty} h(t,u) P(t,u) du
$$

Let us call $V_1$ the value of the portfolio made of the contingent claim, with
value process $C$, financed at the risk-free rate $r(t)$, and $V_2$
that of the replicating portfolio consisting of bonds and of money invested
in the bank account at the risk-free interest rate $r(t)$ (equal to the the spot rate
$f(t,0)$ which may fluctuate but is risk-less in the sense that it
gives the instantaneous payoff of cash invested in the bank account). At time $0 \leq t
\leq T$, the variation of $V_1$ with time, discounted by the risk-free
interest rate, is
\be
d_t V_1 (t)= d_t C(t) - r(t)C(t)dt~.
\label{callll}
\ee
The variation of $V_2$ discounted by the risk-free interest rate is
\be
d_t V_2 (t)= \int_t^{\infty} du~h(t,u) [d_t P(t,u) - r(t) P(t,u)dt]~.
\label{propori}
\ee
There are no other terms in (\ref{propori}) as a sale or purchase of bonds and
deposit or withdrawal on the bank account correspond only to a change of
the nature of the investment but not to a change of wealth. The maturity of
a bond is not a change of wealth either. Only the variations of the bond
prices have to be taken into account. Note that the term $d_t B(t) - r(t) B(t)dt$
vanishes identically from the definition of the spot rate and thus the amount
invested in the bank account does not contribute to the discounted variation of 
$V_2$. The standard replication argument \cite{Merton}
for option pricing and hedging amounts to 
equate the variations in values of the two portfolios. We need some more ingredients
before carrying out this program (see the appendix for a pedagogical exposition of this
formulation for the standard European call option problem). See Ref.\cite{BSor}
for a different approach in the non-Gaussian case.

We assume that bond prices are driven by one of the stochastic string
processes $Z(t,x)$ introduced in \cite{Santa} and used
above. 
$C(t)$ is a priori a function of a continuous infinity of bond prices at
time $t$. The
relevant mathematical tool to calculate $d_t C$ is that of functional
derivation.
We have, up to order $dt$,
$$
d_t C = {\partial C \over \partial t} dt + \int_t^{\infty} du
~{\partial C \over \partial P(t,u)} d_t P(t,u)
$$
$$
+ {1 \over 2} \int_t^{\infty} du
\int_t^{\infty} dv {\partial^2 C \over \partial P(t,u) \partial P(t,v)}
\Cov \left[ d_t P(t,u)~,~ d_t P(t,v) \right]
$$
\be
+ {\partial C \over \partial r} d_t r 
+ {1 \over 2}  {\partial^2 C \over \partial r^2} \Var[d_t r] 
+ {1 \over 2} \int_t^{\infty} du {\partial^2 C \over \partial r \partial P(t,u)} 
\Cov \left[ d_t P(t,u)~,~ d_t r(t) \right
]~.
\label{dcdcdcdcd}
\ee
In (\ref{dcdcdcdcd}), we have taken into account that the option price $C$ is also
a function of the stochastic spot rate $r(t)$.

$d_t P(t,s)$ is given by \cite{Santa}
\be
d_t P(t,s) = dt ~P(t,s)  \biggl[ f(t,0)  - P(t,s) \int_0^{s-t} dy ~\sigma(t,y) ~d_t Z(t,y)\biggl]~.
\label{rppppn}
\ee
This expression can be derived directly from (\ref{erdxwwxx}), (\ref{bphhbcgg}) and Ito's
calculus.
Thus
$$
\Cov \left[ d_t P(t,u),d_t P(t,v) \right] =
$$
\be
P(t,u) ~ P(t,v)
\int_0^{u-t} dy ~\sigma(t,y) \int_0^{v-t} dy' ~\sigma(t,y')
 \Cov \left[ d_t Z(t,y),~ d_t Z(t,y') \right]~.
\label{rppppuuu}
\ee
We also have 
\be
\Var[d_t r] = [\sigma(t,0)]^2 \Var[d_t Z(t,0)] ~,
\ee
and
\be
\Cov \left[ d_t P(t,s),d_t r(t) \right] = - \sigma(t,0)~P(t,s)~
\int_0^{s-t} dy~\sigma(t,y)~\Cov \left[ d_t Z(t,y),~ d_t Z(t,0) \right]~.
\ee

The portfolio of bonds replicates the option if $dV_1(t)=dV_2(t)$.
We already see that a necessary condition for the
 replication to be perfect (the market to be complete) is that
the replicating portfolio is made of a continuous infinity of bonds.
Technically, this comes from the fact that the time differential of the
option price $C$ leads, by the functional differential, to a continuous
integral over all bonds with time-to-maturity larger than $t$ when the
payoff is indeed dependent on the continuous infinity of bond maturities.

The stochastic part proportional to $d_t P(t,u)$ in the replicating equation
$dV_1(t)=dV_2(t)$ cancels out if we choose
\be
h(t,u) = {\partial C(t) \over \partial P(t,u)}~,
\label{deltatlale}
\ee
which is the usual delta hedging. But this is not enough as the stochastic term 
proportional to $d_t r$ still remains. From the fact that $r(t) \equiv f(t,0)$ and
by definition 
\be
P(t,s) = \exp\{-\int_0^{s-t} dy~f(t,y)\}~,
\label{dsfcxcxcxc}
\ee
 we see that bonds close to 
maturity are driven by the same stochastic innovations as the spot rate. To make sense
of this statement, one has to be careful on how the limit $s \to t$ is taken. The standard
way to tackle this problem is to remember that the continuous formulation is nothing but
a limit $\delta t \to 0$ of discrete time increments. We thus have, instead of 
(\ref{dsfcxcxcxc}),
\be
P(t,s) = \exp\{- \delta t ~[f(t,0) + f(t,\delta t) + f(t,2\delta t) +...+ f(t, (n-1) \delta t)]\}~,
\ee
where $n = (s-t)/\delta t$. In particular, 
\be
P(t,t+ \delta t) = \exp\{- \delta t ~f(t,0)\} = 1 - \delta t ~f(t,0)~,
\label{rzedxwcqh}
\ee
where the second equality becomes asymptotically
exact for very small $\delta t$. This shows that it is
possible in principle to hedge the spot rate by bonds that are very close to maturing.
From (\ref{rzedxwcqh}), we obtain
\be
d_t P(t,t+ \delta t) = - \delta t ~d_t f(t,0) = - \delta t ~d_t r(t)~.
\label{rzedkjjjxwcqh}
\ee
From this, we see that 
the stochastic part proportional to $d_t r(t)$ in the replicating equation
$dV_1(t)=dV_2(t)$ cancels out if we add the quantity of bonds
\be
\delta h(t,t+\delta t) = -{1 \over \delta t} {\partial C(t) \over \partial r(t)}~,
\label{delsdfe}
\ee
to the previous quantity $h(t,t+\delta t) = {\partial C(t) \over \partial P(t,t+\delta t)}$
obtained from (\ref{deltatlale}) for the bonds going to maturity.
Notice that the two quantities are equal since one can
replace $- \delta t \partial r(t)$ by $\partial P(t,t+ \delta t) $ as seen from 
(\ref{rzedkjjjxwcqh}). 

To summarize, the hedging strategy is given by 
\be
h(t,u) = [2 - Y(u-\delta t)] ~{\partial C(t) \over \partial P(t,u)}~,
\label{deltauuuyyale}
\ee
where $Y(x)$ is the Heaviside function equal to $1$ for $x\geq 1$ and zero otherwise.
Thus, we recover the usual delta hedging, except for the factor $2$ for bonds close
to maturation which results from the existence of a stochastic spot interest rate. 

The {\it deterministic} equation of the option price is then
$$
{1 \over 2} \int_t^{\infty} du \int_t^{\infty} dv ~A(t,u,v) ~{\partial^2
C(t) \over \partial P(t,u) \partial P(t,v)}  P(t,u)~P(t,v) +
{1 \over 2} \int_t^{\infty} du {\partial^2 C \over \partial r \partial P(t,u)} 
\Cov \left[ d_t P(t,u)~,~ d_t r(t) \right] 
$$
\be
+ {\Var[d_t r]  \over 2}  {\partial^2 C \over \partial r^2} +
{\partial C(t) \over \partial t} - r(t) C(t) +  r(t) \int_t^{\infty}
du~[2 - Y(u-\delta t)]~{\partial C(t) \over \partial P(t,u)}  P(t,u) = 0
\label{finallfinid}
\ee
where
\be
A(t,u,v) \equiv \int_0^{u-t} dy
~\sigma(t,y) \int_0^{v-t} dy' ~\sigma(t,y') c_Z(t,y,y')~.
\label{rdcxcxcpo}
\ee

This equation is correct only when $\sigma(t,x)$ is not 
an explicit function of the forward rates $f$, a case corresponding to 
linear SPDE's (\ref{retfdfqsdqund}). 
When this is not the case, i.e. when $\sigma(t,x)$ is a function of $f$,
we see from (\ref{finallfinid}) with (\ref{rdcxcxcpo}) that
$C$ should also be a function of $f$, in addition to be dependent on
$t,\{P(s)\}$ and $r(t)$. As a consequence, the total time derivative
$d_t C$ must contains terms involving partial derivatives with respect to
$f$ that must be treated self-consistently with the other terms in
(\ref{dcdcdcdcd}).

\subsection{Bond derivatives}

The general case simplifies greatly when the contingent claim has a
payoff that can be written as a function of a single bond price. We see
that pricing and hedging these derivatives is much easier.
This is very interesting since most derivatives that are actually traded have
payoffs that can be written as a function of a finite number of bond prices.

The simplest case of a bond option is a European call or put on a
zero-coupon bond.
Consider a call with maturity $s$ on a bond with maturity $s+\tau$, with
strike price $K$. The payoff at maturity is the amount
\be
C(s, P(s,s+\tau)) = \max(P(s,s+\tau) - K, 0)~,
\label{pricepre}
\ee
The price of this claim at time $t<s$ will only be a function of
$P(t,s+\tau)$,
and the derivative can be hedged with a position in this bond alone.

In the same manner, caps \footnote{A cap guarantees a maximum interest rate for 
borrowing over a determined time horizon.}, floors \footnote{A floor guarantees a
minimum interest rate for an investment over a determined time horizon.},
collars \footnote{A collar is a contract in which 
the buyer is guaranteed an interval of interest rates, with a maximum rate. Its sale
is thus the association of the buy of a cap and the sell of a floor.}
and swaptions \footnote{A swaption is an option on a swap. Simply put,
an interest rate swap is a contract 
 between two parties that exchange for a determined time two interest rates, for instance
 a short-term and a long-term.} have payoffs that, in general, can be written as a function of
the prices of a finite set of bonds. The simplest representative example of
this type of
options is a caplet. A caplet (settled in arrears) pays at
maturity, the maximum of the LIBOR (London Interbank Offered Rate)
rate for the caplet period (set at the
beginning of the period) minus the
cap rate and zero, multiplied by the principal amount. Let the current time
be $t$, the beginning of the caplet period be $s$ and the length of the
caplet period be $\tau$. Denote the
principal amount by $V$, let the LIBOR rate be $L(s,s+\tau)$, and the cap rate
be $K$. Then, the payoff at date $s+\tau$ will be
$$ C(s+\tau) = V \tau \max(L-K,0) $$
that we can express in our notation as
$$
C(s+\tau) = V \max \left( e^{\int_0^{\tau} f(s,y)dy}-1 - K \tau,0 \right)
$$
$$
= V \max \left( \frac{1}{P(s,s+\tau)}-1 - K \tau,0 \right)
$$
Finally, this payoff is known at time $s$, so we can write it as
$$ C(s) = V \max \left( 1 - (1+K\tau)P(s,s+\tau) ,0 \right)$$
Again, the payoff only depends on the price of a single bond, so that the
claim can be priced and hedged with that bond.

If the time $t$ value of the contingent claim depends only on $t$
and $P(t,s)$, the replicating portfolio can have
only the bond $P(t,s)$. Then, the previous calculation simplifies as all
functional derivations transform into the usual derivation and we get,
instead of (\ref{finallfinid})\,:
$$
{1 \over 2} A(t,s) [P(t,s)]^2  {\partial^2 C \over \partial [P(t,s)]^2 }
+ {1 \over 2} {\partial^2 C \over \partial r(t) \partial P(t,s)} 
\Cov \left[ d_t P(t,s)~,~ d_t r(t) \right] 
+ {\Var[d_t r]  \over 2}  {\partial^2 C \over \partial r^2} 
$$
\be
+  r(t) P(t,s) {\partial C \over \partial P(t,s)} 
+ r(t) P(t,t+\delta t) {\partial C \over \partial P(t,t+\delta t)}
+ {\partial C \over \partial t} - r(t) C  = 0
\label{finallughdh}
\ee
where
\be
A(t,s) = \int_0^{s-t} dy ~\sigma(t,y) \int_0^{s-t} dy' ~\sigma(t,y')
~c_Z(t,y,y')~.
\label{aaaaxsssx}
\ee
The equation (\ref{finallughdh}) is similar to the usual Black-Scholes equation
but has two differences. First it has a time-dependent diffusion
coefficient $A(t,x)$. But more importantly, it shows that the option price is
function of {\it two} bond prices, the bond underlying the writing of the option
and the bond just before maturation. We see that the expression (\ref{finallughdh})
has a priori no mathematical sense in the continuous limit but takes full sense when we
replace all derivatives by their discrete differences. This is a novel situation
brought about by the structure of our model defined in terms of correlated but different 
shocks for each maturities as seen from equation (\ref{bphhbcgg}).
In fact, the continuous limit can be retrieved by remarking
that ${\partial C \over \partial P(t,t+\delta t)}$ should be of order $\delta t$\,:
\be
{\partial C \over \partial P(t,t+\delta t)} = \delta t ~ w(t, P(t,s))~.
\ee
Then, the term
$r(t) P(t,t+\delta t) {\partial C \over \partial P(t,t+\delta t)}$
in (\ref{finallughdh}) becomes
$- r(t) ~w(t, P(t,s)) ~P(t,t+\delta t)~ {\partial C \over \partial r(t)}$.
This situation is similar to a boundary layer for singular perturbation
problems and in hydrodynamics where a special treatment
has to be developed close to a boundary, here the spot maturity.

This equation contrasts with the pricing PDE that is usually presented 
in term structure models. In models with state variables such as CIR, the objective is 
to price derivatives as a function of those state variables and time, and so the PDE is 
set with respect to them. The solution of the PDE we present is not a function of state 
variables, which do not exist in our model, but rather is a function of the price of the 
bond underlying the derivative.

It is interesting to compare this pricing equation (\ref{finallughdh}) with
(\ref{aaaaxsssx}) to the corresponding pricing equation for the standard HJM model with
a single Brownian motion driving the full forward rate curve. In this case, we have
\be
d_t P(t,s) = P(t,s) \biggl[ \phi(t) \int_0^{s-t} \sigma(t,y)~dy  \biggl] dt
+ dW(t) ~\int_0^{s-t} dy ~\sigma(t,y)~.
\label{bprigdfvvvx}
\ee
Thus
\be
\Var \left[ d_t P(t,s) \right] = dt~[P(t,s)]^2
~\biggl( \int_0^{s-t} \sigma(t,y)~dy \biggl)^2~.
\ee
Using the same method as above, except for the term 
$r(t) P(t,t+\delta t) {\partial C \over \partial P(t,t+\delta t)}$ which is absent, 
we get the same pricing equation
(\ref{finallughdh}) with
the instantaneous ``diffusion coefficient''
\be
A(t,s)= \biggl( \int_0^{s-t} \sigma(t,y)~dy \biggl)^2 ~,
\label{aadddddsssx}
\ee
instead of (\ref{aaaaxsssx}). The expression (\ref{aaaaxsssx}) reduces to 
(\ref{aadddddsssx}) for $c_Z = 1$ as it should, 
i.e. for perfect correlations along the time-to-maturity
axis, which corresponds to a single Brownian process
(single factor) driving the whole forward rate curve.

\section{Conclusion}

All previous models of interest rates envision the fluctuations of the forward rate
curve as driven by one or several (multi-factors) ``zero-dimensional'' random walk processes 
acting either on the left-end-point of the string,
the short-rate, or on the whole function $f(t,x)$ simultaneously. 
To sum up the situation pictorially, in the first class of models,
the forward rate curve is like a whip held by a shaking hand (the central banks!?), while
in the second class, 
imagine a large hammer of the size of the full curve hitting it simultaneously
through an irregular pillow (the volatility curve) transmitting 
inhomogeneously the impact along the curve.

The present paper, which extends Ref.\cite{Santa}, explores the more general
situation where all the points of the forward curve are simultaneously driven by
random shocks. Pictorially, this is like a string submitted to the incessant impacts of rain
drops. We have argued that the condition of absence of arbitrage opportunity
provides a constructive principle for the establishment of a general theory of 
interest rate dynamics. We have derived the general condition (\ref{terfdvxc})  that a stochastic
partial differential equation (SPDE) for the forward rate must obey. Using our previous
parametrization in terms of string shocks \cite{Santa}, we have derived the 
general structure (\ref{retfdfqsdqund}) of the SPDE for forward rates under the no-arbitrage
condition.  It is noteworthy that the 
no-arbitrage condition excludes the parabolic class of partial differential equations, 
i.e. those that usually describe the physical strings submitted to strong fluctuations! 

In a second part, we have derived a general approach to price and hedge options 
defined on bonds and thus on the forward rate curve, in terms of functional
partial differential equations. We have found that the usual Black-Scholes replication
strategy can be adapted to this situation, provided that special terms be added to
account self-consistently for the instantaneous interest rate. 
Our approach is limited to the case where the volatility 
$\sigma(t,x)$ does not depend on $f(t,x)$ itself, thus excluding the a priori most
interesting class of stochastic {\it non-linear}
partial differential equations. 

Let us end with a conjecture. It may be conceivable that
a simple form for the nonlinear dependence of $\sigma$ as a function of
$f(t,x)$, with one or two real parameters, might embody the full phenomenology of observed
forward rate statistics, hence constituting a truly fundamental theory of forward rate curves.
It would be fundamental in the sense that the properties of the shocks would be generated 
dynamically by the nonlinear interactions similarly for instance
to the coherent vortices in turbulence for
instance, and in contrast to the usual external shock assignments in the existing linear models.
The complex behavior would then result from the interplay between 
external factors represented by the noise source and the non-linear
dynamics. Persuing along this conjectural tone, this approach might provide an line of 
attack for understanding the recent empirical 
finding \cite{Causal} of a causal information cascade across 
scales in volatilities, occurring
 from large time scales to short times scales in a way very similar to
the celebrated Kolmogorov energy cascade proposed for fluid turbulence.

\vskip 1cm
{\bf Acknowlegments}\,: 
I am very grateful to M. Brennan
and especially P. Santa-Clara for helpful discussions and to D. Stauffer for 
a careful reading of the manuscript.

\pagebreak
APPENDIX

The simplest option pricing problem (the so called 
`European call options')
is the following\,: suppose that an operator wants to buy from a bank a given share,
a certain time
$t=T$ from now ($t=0$), at a fixed `striking' price $x_c$. If the share value at
$t=T$, $x(T)$, exceeds $x_c$, the operator `exercises' his option. His  
gain, when reselling 
immediately at the current price $x(T)$, is thus the
difference $x(T)-x_c$. On the contrary, if $x(T) < x_c$ the operator does
 not buy the share.  
What is the price ${\cal C}$ of this option, and what trading strategy $\phi$ should be
followed by the bank between now and $T$, depending on what the share value $x(t)$
actually does between $t=0$ and $t=T$? 

In the standard treatment in continuous time \cite{Merton}, one
forms the portfolio $F=-{\cal C} + x\phi$ such that $d_t F$ remains
indentically zero. Here, we propose a slightly different derivation
of Black-Scholes' result based on the idea that the bank constructs a portfolio that
replicates the option exactly, thereby eliminating all risks. This is of course only
valid under the restricted assumptions of continuous trading, Gaussian random walk
of market prices, and absence of market imperfections and transaction costs.

The idea is to compare the two points of view of the option buyer and of the option
seller. During the time increment $dt$, their respective change of wealth discounted
by the risk-free interest rate $r$ is 
\be
dW_a = d{\cal C} - r {\cal C} dt~,
\ee
for the buyer who owns only the option and
\be
dW_v = \phi [dx - rx dt]~,
\ee
for the seller who possesses a portfolio made of $\phi$ underlying stock shares.
$d{\cal C} - r {\cal C} dt$ is the gain or loss of the buyer above the risk-free
return. $\phi [dx - rx dt]$ is the gain or loss of the seller for a price variation
of the stock above the risk-free return. The fair price of the option 
and the hedging strategy
that the seller must follow are those such that the return is the same for both 
traders, namely
\be
dW_a = dW_v~.
\label{rtfgvbb}
\ee
This equality makes concrete the fundamental idea of Black and Scholes that
the fair price and the hedging strategy are univoquely determined for all
traders, independently of their risk aversion, if the seller can replicate the option.

Using Ito's formula to calculate $d{\cal C}$, we have\,:
\be
d{\cal C} = \frac{\partial {\cal C}(x,x_c,T-t)}{\partial t} dt + 
 \frac{\partial {\cal C}(x,x_c,T-t)}{\partial x} dx +  \frac{D}{2}
 \frac{\partial^2 {\cal C}(x,x_c,T-t)}{\partial x^2} dt  .
\ee
Inserting this expression in (\ref{rtfgvbb}) shows immediately that, if one
makes the choice
\be
\phi = \phi^* \equiv \frac{\partial {\cal C}(x,x_c,T-t)}{\partial x}~,
\label{rezaqsxwcc}
\ee
then the only stochastic term, namely $dx$, cancels out!
The comparison between the two returns of the buyer and seller become certain.
In this case, the equation (\ref{rtfgvbb}) provides the following deterministic
partial differential equation for $\cal C$: 
\be 
\frac{\partial {\cal C}(x,x_c,T-t)}{\partial t} + r x  \frac{\partial
{\cal C}(x,x_c,T-t)}{\partial x} 
 + \frac{D}{2} \frac{\partial^2 {\cal C}(x,x_c,T-t)}{\partial x^2}  - r {\cal
C}(x,x_c,T-t) = 0 ~,
\ee 
with bouncary conditions the value of the option at maturity
$C(x,x_c,0) \equiv \max(x-x_c,0)$. The solution of this equation is the celebrated
Black and Scholes formula \cite{Black}. The hedging strategy is then the 
derivative of this solution with respect to $x$. This derivation shows
very straightforwardly why the average return of the underlying stock does 
not appear\,; it has been avoided by the replication condition (\ref{rtfgvbb})
with (\ref{rezaqsxwcc}).

\pagebreak

\end{document}